\documentclass[12pt]{revtex4}    
\usepackage{amssymb,epsf}
\usepackage{latexsym}

\begin{document}

\title{Magnetic dilaton strings in anti-de Sitter spaces}
\author{Ahmad Sheykhi \footnote{sheykhi@mail.uk.ac.ir}}
\address{Department of Physics, Shahid Bahonar University, P.O. Box 76175, Kerman, Iran\\
         Research Institute for Astronomy and Astrophysics of Maragha (RIAAM), Maragha, Iran}

\begin{abstract}
With an appropriate combination of three Liouville-type dilaton
potentials, we construct a new class of spinning magnetic dilaton
string solutions which produces a longitudinal magnetic field in
the background of anti-de Sitter spacetime. These solutions have
no curvature singularity and no horizon, but have a conic
geometry. We find that the spinning string has a net electric
charge which is proportional to the rotation parameter. We present
the suitable counterterm which removes the divergences of the
action in the presence of dilaton potential. We also calculate the
conserved quantities of the solutions by using the counterterm
method.

\end{abstract}

\maketitle
\section{Introduction\label{Intro}}
The construction and analysis of black hole solutions in the
background of anti-de Sitter (AdS) spaces is a subject of much
recent interest. This interest is primarily motivated by the
correspondence between the gravitating fields in an AdS spacetime
and conformal field theory living on the boundary of the AdS
spacetime \cite{Witt1}. This equivalence enables one to remove the
divergences of the action and conserved quantities of gravity in
the same way as one does in field theory. It was argued that the
thermodynamics of black holes in AdS spaces can be identified with
that of a certain dual conformal field theory (CFT) in the high
temperature limit \cite{Witt2}. Having the AdS/CFT correspondence
idea at hand, one can gain some insights into thermodynamic
properties and phase structures of strong 't Hooft coupling
conformal field theories by studying the thermodynamics of
asymptotically AdS black holes.

On another front, scalar coupled black hole solutions with
different asymptotic spacetime structure is a subject of interest
for a long time. There has been a renewed interest in such studies
ever since new black hole solutions have been found in the context
of string theory. The low energy effective action of string theory
contains two massless scalars namely dilaton and axion. The
dilaton field couples in a nontrivial way to other fields such as
gauge fields and results into interesting solutions for the
background spacetime. It was argued that with the exception of a
pure cosmological constant, no dilaton-de Sitter or anti-de Sitter
black hole solution exists with the presence of only one
Liouville-type dilaton potential \cite{MW}. Recently, the dilaton
potential leading to (anti)-de Sitter-like solutions of dilaton
gravity has been found \cite{Gao1}. It was shown that the
cosmological constant is coupled to the dilaton in a very
nontrivial way. With the combination of three Liouville-type
dilaton potentials, a class of static dilaton black hole solutions
in (A)dS spaces has been obtained by using a coordinates
transformation which recast the solution in the schwarzschild
coordinates system \cite{Gao1}. More recently, a class of charged
rotating dilaton black string solutions in four-dimensional
anti-de Sitter spacetime has been found in \cite{shey1}. Other
studies on the dilaton black hole solutions in (A)dS spaces have
been carried out in \cite{Gao2,shey2}.

In this Letter, we turn to the investigation of asymptotically AdS
spacetimes generated by static and spinning string sources in
four-dimensional Einstein-Maxwell-dilaton theory which are
horinzonless and have nontrivial external solutions. The
motivation for studying such kinds of solutions is that they may
be interpreted as cosmic strings. Cosmic strings are topological
structure that arise from the possible phase transitions to which
the universe might have been subjected to and may play an
important role in the formation of primordial structures. A short
review of papers treating this subject follows. The
four-dimensional horizonless solutions of Einstein gravity have
been explored in \cite{Vil,Ban}. These horizonless solutions
\cite{Vil,Ban} have a conical geometry; they are everywhere flat
except at the location of the line source. The spacetime can be
obtained from the flat spacetime by cutting out a wedge and
identifying its edges. The wedge has an opening angle which turns
to be proportional to the source mass. The extension to include
the Maxwell field has also been done \cite{Bon}. Static and
spinning magnetic sources in three and four-dimensional
Einstein-Maxwell gravity with negative cosmological constant have
been explored in \cite{Lem1,Lem2}. The generalization of these
asymptotically AdS magnetic rotating solutions to higher
dimensions has also been done \cite{Deh2}. In the context of
electromagnetic cosmic string, it has been shown that there are
cosmic strings, known as superconducting cosmic strings, that
behave as superconductors and have interesting interactions with
astrophysical magnetic fields \cite{Wit2}. The properties of these
superconducting cosmic strings have been investigated in
\cite{Moss}. It is also of great interest to generalize the study
to the  dilaton gravity theory \cite{Fer}. While exact magnetic
rotating dilaton solution in three dimensions has been obtained in
\cite{Dia}, two classes of magnetic rotating solutions in four
\cite{Deh4} and higher dimensional dilaton gravity in the presence
of one Liouville-type potential have been constructed \cite{SDR}.
Unfortunately, these solutions \cite{Deh4,SDR} are neither
asymptotically flat nor (A)dS. The purpose of the present Letter
is to construct a new class of static and spinning magnetic
dilaton string solutions which produces a longitudinal magnetic
field in the background of anti-de Sitter spacetime. We will also
present the suitable counterterm which removes the divergences of
the action, and calculate the conserved quantities by using the
counterterm method.

\section{Basic Equations\label{Basic}}
Our starting point is the four-dimensional
Einstein-Maxwell-dilaton action
\begin{eqnarray}
I_{G} &=&-\frac{1}{16\pi }\int_{\mathcal{M}}d^{4}x\sqrt{-g}\left(
R\text{ }-2\partial_{\mu}\Phi \partial^{\mu}\Phi-V(\Phi
)-e^{-2\alpha \Phi
}F_{\mu \nu }F^{\mu \nu }\right)   \nonumber \\
&&-\frac{1}{8\pi }\int_{\partial \mathcal{M}}d^{3}x\sqrt{-\gamma
}\Theta (\gamma ),  \label{Act}
\end{eqnarray}
where ${R}$ is the scalar curvature, $\Phi$ is the dilaton field,
$F_{\mu \nu }=\partial _{\mu }A_{\nu }-\partial _{\nu }A_{\mu }$
is the electromagnetic field tensor, and $A_{\mu }$ is the
electromagnetic potential. $\alpha $ is an arbitrary constant
governing the strength of the coupling between the dilaton and the
Maxwell field. The last term in Eq. (\ref{Act}) is the
Gibbons-Hawking surface term. It is required for the variational
principle to be well-defined. The factor $\Theta$ represents the
trace of the extrinsic curvature for the boundary ${\partial
\mathcal{M}}$ and $\gamma$ is the induced metric on the boundary.
While $\alpha=0$ corresponds to the usual Einstein-Maxwell-scalar
theory, $\alpha=1$ indicates the dilaton-electromagnetic coupling
that appears in the low energy string action in Einstein's frame.
For arbitrary value of $\alpha $ in  AdS space the form of the
dilaton potential $V(\Phi)$ is chosen as \cite{Gao1}
\begin{equation}
V(\Phi )=\frac{2\Lambda}{3(\alpha^2+1)^{2}}\left[ {\alpha}^{2}
\left( 3\,{\alpha}^{2}-1 \right) {e ^{-2\Phi/\alpha}}+ \left(
3-{\alpha}^{2} \right) {e^{2\,
\alpha\,\Phi}}+8\,{\alpha}^{2}{e^{\Phi(\alpha-1/\alpha)}} \right]
. \label{v1}
\end{equation}
Here $\Lambda $ is the cosmological constant. It is clear that the
cosmological constant is coupled to the dilaton field in a very
nontrivial way. This type of the dilaton potential was introduced
for the first time by Gao and Zhang \cite{Gao1}. They derived, by
applying a coordinates transformation which recast the solution in
the Schwarzchild coordinates system, the static dilaton black hole
solutions in the background of (A)dS universe. For this purpose,
they required the existence of the (A)dS dilaton black hole
solutions and extracted successfully the form of the dilaton
potential leading to (A)dS-like solutions. They also argued that
this type of derived potential can be obtained when a higher
dimensional theory is compactified to four dimensions, including
various supergravity models \cite{Gid}. In the absence of the
dilaton field the action (\ref{Act}) reduces to the action of
Einstein-Maxwell gravity with cosmological constant. Varying the
action (\ref{Act}) with respect to the gravitational field $g_{\mu
\nu }$, the dilaton field $\Phi $ and the gauge field $A_{\mu }$,
yields
\begin{equation}
{R}_{\mu \nu }=2 \partial _{\mu }\Phi \partial _{\nu }\Phi
+\frac{1}{2}g_{\mu \nu }V(\Phi )+2e^{-2\alpha \Phi}\left( F_{\mu
\eta }F_{\nu }^{\text{ }\eta }-\frac{1}{4}g_{\mu \nu }F_{\lambda
\eta }F^{\lambda \eta }\right) ,  \label{FE1}
\end{equation}
\begin{equation}
\nabla ^{2}\Phi =\frac{1}{4}\frac{\partial V}{\partial \Phi
}-\frac{\alpha }{2}e^{-2\alpha \Phi}F_{\lambda \eta }F^{\lambda
\eta }, \label{FE2}
\end{equation}
\begin{equation}
\partial _{\mu }\left( \sqrt{-g}e^{-2\alpha \Phi}F^{\mu \nu
}\right) =0.  \label{FE3}
\end{equation}
The conserved mass and angular momentum of the solutions of the
above field equations can be calculated through the use of the
substraction method of Brown and York \cite{BY}. Such a procedure
causes the resulting physical quantities to depend on the choice
of reference background. A well-known method of dealing with this
divergence for asymptotically AdS solutions of Einstein gravity is
through the use of counterterm method inspired by AdS/CFT
correspondence \cite{Mal}. In this Letter, we deal with the
spacetimes with zero curvature boundary, $R_{abcd}(\gamma )=0$,
and therefore the counterterm for the stress energy tensor should
be proportional to $\gamma ^{ab}$. We find the suitable
counterterm which removes the divergences of the action in the
form (see also \cite{otha})
\begin{equation}\label{cont}
 I_{ct}=-\frac{1}{8\pi }\int_{\partial \mathcal{M}}d^{3}x\sqrt{-\gamma
}\left(-\frac{1}{l}+\frac{\sqrt{-6V(\Phi)}}{2} \right).
\end{equation}
One may note that in the absence of a dilaton field where we have
$V(\Phi)=2\Lambda=-6/l^2$, the above counterterm has the same form
as in the case of asymptotically AdS solutions with zero-curvature
boundary. Having the total finite action $I =
I_{G}+I_{\mathrm{ct}}$ at hand, one can use the quasilocal
definition  to construct a divergence free stress-energy tensor
\cite{BY}. Thus the finite stress-energy tensor in
four-dimensional Einstein-dilaton gravity with three
Liouville-type dilaton potentials (\ref{v1}) can be written as
\begin{equation}
T^{ab}=\frac{1}{8\pi }\left[ \Theta ^{ab}-\Theta \gamma
^{ab}+\left(-\frac{1}{l}+\frac{\sqrt{-6V(\Phi)}}{2} \right)\gamma
^{ab}\right]. \label{Stres}
\end{equation}
The first two terms in Eq. (\ref{Stres}) are the variation of the
action (\ref{Act}) with respect to $\gamma _{ab}$, and the last
two terms are the variation of the boundary counterterm
(\ref{cont}) with respect to $\gamma _{ab}$. To compute the
conserved charges of the spacetime, one should choose a spacelike
surface $ \mathcal{B}$ in $\partial \mathcal{M}$ with metric
$\sigma _{ij}$, and write the boundary metric in ADM
(Arnowitt-Deser-Misner) form:
\[
\gamma _{ab}dx^{a}dx^{b}=-N^{2}dt^{2}+\sigma _{ij}\left( d\varphi
^{i}+V^{i}dt\right) \left( d\varphi ^{j}+V^{j}dt\right) ,
\]
where the coordinates $\varphi ^{i}$ are the angular variables
parameterizing the hypersurface of constant $r$ around the origin,
and $N$ and $V^{i}$ are the lapse and shift functions,
respectively. When there is a Killing vector field $\mathcal{\xi
}$ on the boundary, then the quasilocal conserved quantities
associated with the stress tensors of Eq. (\ref{Stres}) can be
written as
\begin{equation}
Q(\mathcal{\xi )}=\int_{\mathcal{B}}d^{2}x \sqrt{\sigma }T_{ab}n^{a}%
\mathcal{\xi }^{b},  \label{charge}
\end{equation}
where $\sigma $ is the determinant of the metric $\sigma _{ij}$, $\mathcal{%
\xi }$ and $n^{a}$ are, respectively, the Killing vector field and
the unit normal vector on the boundary $\mathcal{B}$. For
boundaries with timelike ($\xi =\partial /\partial t$) and
rotational ($\varsigma =\partial /\partial \phi $) Killing vector
fields, one obtains the quasilocal mass and angular momentum
\begin{eqnarray}
M &=&\int_{\mathcal{B}}d^{2}x \sqrt{\sigma }T_{ab}n^{a}\xi ^{b},
\label{Mastot} \\
J &=&\int_{\mathcal{B}}d^{2}x \sqrt{\sigma }T_{ab}n^{a}\varsigma
^{b}.  \label{Angtot}
\end{eqnarray}
These quantities are, respectively, the conserved mass and angular
momenta of the system enclosed by the boundary $\mathcal{B}$. Note
that they will both depend on the location of the boundary
$\mathcal{B}$ in the spacetime,
although each is independent of the particular choice of foliation $\mathcal{%
B}$ within the surface $\partial \mathcal{M}$.

\section{Static magnetic dilaton string \label{static}}

Here we want to obtain the four-dimensional solution of Eqs.
(\ref{FE1})-(\ref{FE3}) which produces a longitudinal magnetic
fields along the $z$ direction. We assume the following form for
the metric \cite{Lem1}
\begin{equation}
ds^{2}=-\frac{\rho ^{2}}{l^{2}}R^{2}(\rho )dt^{2}+\frac{d\rho ^{2}}{f(\rho )}%
+l^{2}f(\rho )d\phi ^{2}+\frac{\rho ^{2}}{l^{2}}R^{2}(\rho
)dz^{2}. \label{Met1}
\end{equation}
The functions $f(\rho)$ and $R(\rho)$ should be determined and $l$
has the dimension of length which is related to the cosmological
constant $\Lambda $ by the relation $l^{2}=-3/\Lambda$. The
coordinate $z$ has the dimension of length and ranges  $-\infty
<z<\infty$, while the angular coordinate $\phi $ is dimensionless
as usual
and ranges $0\leq \phi <2\pi $. The motivation for this curious choice of the metric gauge $%
[g_{tt}\varpropto -\rho ^{2}$ and $(g_{\rho \rho })^{-1}\varpropto
g_{\phi \phi }]$ instead of the usual Schwarzschild gauge
$[(g_{\rho \rho })^{-1}\varpropto g_{tt}$ and $g_{\phi \phi
}\varpropto \rho ^{2}] $ comes from the fact that we are looking
for a magnetic solution instead of an electric one. It is
well-known that the electric field is associated with the time
component, $A_t$, of the vector potential while the magnetic field
is associated with the angular component $A_{\phi}$. From the
above fact, one can expect that a magnetic solution can be written
in a metric gauge in which the components $g_{tt}$ and $g_{\phi
\phi}$ interchange their roles relatively to those present in the
Schwarzschild gauge used to describe electric solutions
\cite{Lem1}. The Maxwell equation (\ref{FE3}) can be integrated
immediately to give
\begin{equation}  \label{Ftr}
F_{\phi \rho}=\frac{ql e^{2\alpha \Phi }}{\rho ^2 R^{2}},
\end{equation}
where $q$, an integration constant, is the charge parameter which
is related to the electric charge of the rotating string, as will
be shown below. Inserting the Maxwell fields (\ref{Ftr}) and the
metric (\ref{Met1}) in the field equations (\ref{FE1}) and
(\ref{FE2}), we can simplify these equations as
\begin{eqnarray}\label{Eqtt}
&& 2\rho^3 R ^4 f'+ 2\rho^4 R ^3 f' R'+2\rho^2 R ^4 f+8\rho^3 R ^3
f R'+2\rho^4 R ^2 f R'^2 \nonumber \\ &&+2\rho^4 R^3 f R''+ \rho^4
R ^4 V \left( \Phi \right) -2q^2 e^{2\alpha\Phi}=0,
\\&&2\rho^3 R ^4 f'+ \rho^4 R ^4 f'' +8\rho^3 R ^3 f R'+4\rho^4 R ^3
f R''+2\rho^4 R ^3 R' f' \nonumber \\ &&+4 \rho^4 R ^4 f \Phi'^2+
\rho^4 R ^4 V \left( \Phi \right) +2q^2 e^{2\alpha\Phi}=0,
\label{Eqrr}
 \\ &&
2\rho^4 R ^3 R' f'+ \rho^4 R ^4 f'' +2\rho^3 R
^4 f'+\rho^4 R ^4 V
 ( \Phi)+2q^2 e^{2\alpha\Phi}=0,\label{Eqpp}
 \\
&&\rho^4 R ^4 {\Phi'}{f'}+\rho^4 R ^4 {\Phi''} {f}+2\rho^3 R ^4
\Phi' f+2 \rho^4 R ^3 R' {\Phi'}{f}- \rho^4 R ^4
\frac{\partial{V}}{4\partial{\Phi}}+\alpha q^2
e^{2\alpha\Phi}=0,\label{Eq3}
\end{eqnarray}
where the ``prime'' denotes differentiation with respect to
$\rho$. Subtracting Eq. (\ref{Eqpp}) from Eq. (\ref{Eqrr}) we get
\begin{eqnarray}\label{Eqtt-Eqrr}
2R' +\rho R'' + \rho R \Phi'^2=0.
\end{eqnarray}
Then we make the ansatz \cite{shey1}
\begin{equation}\label{Rphi}
 R(\rho)=e^{\alpha \Phi}.
\end{equation}
Substituting this ansatz in Eq. (\ref{Eqtt-Eqrr}), it reduces to
\begin{eqnarray}\label{Eqphi}
\rho \alpha \Phi''+2\alpha \Phi'+\rho(1+{\alpha}^{2}) \Phi'^2=0,
\end{eqnarray}
which has a solution of the form
\begin{equation}
\Phi (\rho)=\frac{\alpha }{\alpha ^{2}+1}\ln (1-\frac{b}{\rho}),
\label{phi}
\end{equation}
where $b$ is a constant of integration related to the mass of the
string, as will be shown. Inserting (\ref{phi}), the ansatz
(\ref{Rphi}), and the dilaton potential (\ref{v1}) into the field
equations (\ref{Eqtt})-(\ref{Eq3}), one can show that these
equations have the following solution
\begin{equation}\label{f(r)}
f(\rho)=\frac{c}{\rho} \left( 1-{\frac {b}{\rho}} \right) ^{{\frac
{1-{\alpha}^{2}}{1+{\alpha}^ {2}}}}-\frac{\Lambda}{3}\,{\rho}^{2}
\left( 1-{\frac {b}{\rho}}
 \right) ^{{\frac {2{\alpha}^{2}}{{\alpha}^{2}+1}}},
\end{equation}
where $c$ is an integration constant. The two constants $c$ and
$b$ are related to the charge parameter via $q^2(1+\alpha^2)=bc$.
It is apparent that this spacetime is asymptotically AdS. In the
absence of a nontrivial dilaton ($\alpha=0$), the solution reduces
to the asymtotically AdS horizonless magnetic string for $\Lambda=
-3/l^2$ \cite{Lem2}.

Then we study the general structure of the solution. It is easy to
show that the Kretschmann scalar $R_{\mu \nu \lambda \kappa
}R^{\mu \nu \lambda \kappa }$ diverges at $\rho =0$ and therefore
one might think that there is a curvature singularity located at
$\rho =0$. However, as we will see below, the spacetime will never
achieve $\rho =0$. Second, we look for the existence of horizons
and, in particular, we look for the possible presence of
magnetically charged black hole solutions. The surface $r = b$ is
a curvature singularity for $\alpha\neq0$. The horizons, if any
exist, are given by the zeros of the function $f(\rho )=(g_{\rho
\rho })^{-1}$. Let us denote the largest positive root of $f(\rho
)=0$ by $r_{+}$. The function $f(\rho )$ is negative for $\rho
<r_{+}$, and therefore one may think that the hypersurface of
constant time and $\rho =r_{+}$ is the horizon. However, the above
analysis is wrong. Indeed, we first notice that $ g_{\rho \rho }$
and $g_{\phi \phi }$ are related by $f(\rho )=g_{\rho \rho
}^{-1}=l^{-2}g_{\phi \phi }$, and therefore when $g_{\rho \rho }$
becomes negative (which occurs for $\rho <r_{+}$) so does $g_{\phi
\phi }$. This leads to an apparent change of signature of the
metric from $+2$ to $-2$. This indicates that we are using an
incorrect extension. To get rid of this incorrect extension, we
introduce the new radial coordinate $r$ as
\begin{equation}
r^{2}=\rho ^{2}-r_{+}^{2}\Rightarrow d\rho ^{2}=\frac{r^{2}}{r^{2}+r_{+}^{2}}%
dr^{2}.  \label{Tr1}
\end{equation}
With this coordinate change, the metric (\ref{Met1}) is
\begin{eqnarray}
ds^{2}
&=&-\frac{r^{2}+r_{+}^{2}}{l^{2}}R^{2}(r)dt^{2}+l^{2}f(r)d\phi
^{2}
\nonumber \\
&&+\frac{r^{2}}{(r^{2}+r_{+}^{2})f(r)}dr^{2}+\frac{r^{2}+r_{+}^{2}}{l^{2}}%
R^{2}(r)dz^{2},  \label{Met2}
\end{eqnarray}
where the coordinates $r$ assumes the values $0\leq r<\infty $,
and $f(r)$, $R(r)$, and $\Phi (r)$ are now given as
\begin{equation}\label{f2}
f(r)=\frac{c}{\sqrt{r^{2}+r_{+}^{2}}} \left( 1-{\frac
{b}{\sqrt{r^{2}+r_{+}^{2}}}} \right) ^{{\frac
{1-{\alpha}^{2}}{1+{\alpha}^
{2}}}}-\frac{\Lambda}{3}\,({r^{2}+r_{+}^{2}}) \left( 1-{\frac
{b}{\sqrt{r^{2}+r_{+}^{2}}}}
 \right) ^{{\frac {2{\alpha}^{2}}{{\alpha}^{2}+1}}},
\end{equation}
\begin{equation}
R(r)=\left(1- \frac{b}{%
\sqrt{r^{2}+r_{+}^{2}}}\right)^{\frac{\alpha^2 }{1+\alpha ^{2}}}, \hspace{%
0.7cm}
\Phi (r)=\frac{\alpha }{1+\alpha ^{2}}\ln \left(1- \frac{b}{%
\sqrt{r^{2}+r_{+}^{2}}}\right). \label{R2}
\end{equation}
One can easily show that the Kretschmann scalar does not diverge
in the range $0\leq r<\infty $. However, the spacetime has a conic
geometry and has a conical singularity at $r=0$, since:
\begin{equation}
\lim_{r\rightarrow 0}\frac{1}{r}\sqrt{\frac{g_{\phi \phi
}}{g_{rr}}}\neq 1. \label{limit}
\end{equation}
That is, as the radius $r$ tends to zero, the limit of the ratio
``circumference/radius'' is not $2\pi $ and therefore the
spacetime has a conical singularity at $r=0$. The canonical
singularity can be removed if one identifies the coordinate $\phi$
with the period
\begin{equation}
\textrm{Period}_{\phi}=2 \pi \left(\lim_{r\rightarrow
0}\frac{1}{r}\sqrt{\frac{g_{\phi \phi }}{g_{rr}}} \right)^{-1}=2
\pi (1-4 \mu), \label{period}
\end{equation}
where
\begin{eqnarray}\label{mu}
1-4\mu &=&\Bigg{\{} \frac{\Lambda l(r_{+}-b)^{{\frac
{2{\alpha}^{2}}{{\alpha}
^{2}+1}}}\left[r_{+}(\alpha^2+1)-b\right]}{3(\alpha^2+1)(b-r_{+})r_{+}^{\frac{\alpha^2-1}{1+\alpha^2}}}
+\frac{l c
(r_{+}-b)^{\frac{1-\alpha^2}{1+\alpha^2}}\left[r_{+}(\alpha^2+1)-2b\right]}{2(\alpha^2+1)(b-r_{+})r_{+}^{\alpha^2+3}}
\Bigg {\}}^{-1}.
\end{eqnarray}
The above analysis shows that near the origin $r=0$, the metric
(\ref{Met2}) describes a spacetime which is locally flat but has a
conical singularity at $r=0$ with a deficit angle $\delta \phi=8
\pi \mu$. Since near the origin the metric (\ref{Met2}) is
identical to the spacetime generated by a cosmic string, by using
the Vilenkin procedure, one can show that $\mu$ in Eq. (\ref{mu})
can be interpreted as the mass per unit length of the string
\cite{Vil2}.

\section{Spinning magnetic dilaton string} \label{Angul}
Now, we would like to endow the spacetime solution (\ref{Met1})
with a rotation. In order to add an angular momentum to the
spacetime, we perform the following rotation boost in the $t-\phi
$ plane
\begin{equation}
t\mapsto \Xi t-a\phi ,\hspace{0.5cm}\phi \mapsto \Xi \phi -\frac{a}{%
l^{2}}t,  \label{Tr}
\end{equation}
where $a$ is a rotation parameter and $\Xi =\sqrt{1+a^{2}/l^{2}}$.
Substituting Eq. (\ref{Tr}) into Eq. (\ref{Met2}) we obtain
\begin{eqnarray}
ds^{2} &=&-\frac{r^{2}+r_{+}^{2}}{l^{2}}R^2(r)\left( \Xi dt-ad\phi
\right) ^{2}+\frac{r^{2}dr^{2}}{(r^{2}+r_{+}^{2})f(r)}
\nonumber \\
&&+l^{2}f(r)\left( \frac{a}{l^{2}}dt-\Xi d\phi \right) ^{2}+\frac{%
r^{2}+r_{+}^{2}}{l^{2}}R^2(r)dz^{2},  \label{Met3}
\end{eqnarray}
where $f(r)$ and $R(r)$ are given in Eqs. (\ref{f2}) and
(\ref{R2}). The non-vanishing electromagnetic field components
become
\begin{eqnarray}
F_{\phi r }=\frac{q\Xi l} {r^{2}+r_{+}^{2} },\hspace{0.7cm}  F_{tr
}=-\frac{a}{\Xi l^{2}}F_{\phi r }.
\end{eqnarray}
The transformation (\ref{Tr}) generates a new metric, because it
is not a permitted global coordinate transformation. This
transformation can be done locally but not globally. Therefore,
the metrics (\ref{Met2}) and (\ref{Met3}) can be locally mapped
into each other but not globally, and so they are distinct. Note
that this spacetime has no horizon and curvature singularity.
However, it has a conical singularity at $r=0$. It is notable to
mention that for $\alpha=0$, this solution reduces to the
asymtotically AdS  magnetic rotating string solution presented in
\cite{Lem2}.

The mass and angular momentum per unit length of the string when
the boundary $ \mathcal{B}$ goes to infinity can be calculated
through the use of Eqs. (\ref{Mastot}) and (\ref{Angtot}). We
obtain
\begin{eqnarray}\label{M}
{M}&=&\frac{\alpha^2(\alpha^2-1)b^{3}}{24\pi l^3(\alpha
^{2}+1)^3}+ \frac{(3\Xi ^{2}-2)c}{16\pi l},
\end{eqnarray}
\begin{eqnarray}
{J}&=&\frac{3\Xi c \sqrt{\Xi^2-1}}{16\pi}.  \label{J}
\end{eqnarray}
For $a=0$ ($\Xi =1$), the angular momentum per unit length
vanishes, and therefore $a$ is the rotational parameter of the
spacetime.

Finally, we compute the electric charge of the solutions. To
determine the electric field one should consider the projections
of the electromagnetic field tensor on special hypersurface. The
normal vectors to such hypersurface for the spacetime with a
longitudinal magnetic field are
\[
u^{0}=\frac{1}{N},\text{ \ }u^{r}=0,\text{ \ }u^{i}=-\frac{V^{i}}{N},
\]
and the electric field is $E^{\mu }=g^{\mu \rho }e^{-2\alpha \Phi
}F_{\rho \nu }u^{\nu }$. Then the electric charge per unit length ${Q%
}$ can be found by calculating the flux of the electric field at infinity,
yielding
\begin{equation}
{Q}=\frac{q\sqrt{\Xi ^{2}-1}}{4 \pi l}.  \label{chden}
\end{equation}
It is worth noting that the electric charge is proportional to the
rotation parameter, and is zero for the case of static solution.
This result is expected since now, besides the magnetic field
along the $\phi $ coordinate, there is also a radial electric field ($%
F_{tr}\neq 0$).\ To give a physical interpretation for the
appearance of the net electric charge, we first consider the
static spacetime. The magnetic field source can be interpreted as
composed of equal positive and negative charge densities, where
one of the charge density is at rest and the other one is
spinning. Clearly, this system produce no electric field since the
net electric charge density is zero, and the magnetic field is
produced by the rotating electric charge density. Now, we consider
the rotating  solution. From the point of view of an observer at
rest relative to the source ($S$), the two charge densities are
equal, while from the point of view of an observe $S^{\prime }$
that follows the intrinsic rotation of the spacetime, the positive
and negative charge densities are not equal, and therefore the net
electric charge of the spacetime is not zero.
\section{Conclusion and discussion}
In conclusion, with an appropriate combination of three
Liouville-type dilaton potentials, we constructed a class of
four-dimensionl magnetic dilaton string solutions which produces a
longitudinal magnetic field in the background of anti-de Sitter
universe. These solutions have no curvature singularity and no
horizon, but have conic singularity at $r=0$. In fact, we showed
that near the origin $r=0$, the metric (\ref{Met2}) describes a
spacetime which is locally flat but has a conical singularity at
$r=0$ with a deficit angle $\delta \phi=8 \pi \mu$, where $\mu$
can be interpreted as the mass per unit length of the string. In
these static spacetimes, the electric field vanishes and therefore
the string has no net electric charge. Then we added an angular
momentum to the spacetime by performing a rotation boost in the
$t-\phi $ plane. For the spinning string, when the rotation
parameter is nonzero, the string has a net electric charge which
is proportional to the magnitude of the rotation parameter. We
found the suitable counterterm which removes the divergences of
the action in the presence of three Liouville-type dilaton
potentials. We also computed the conserved quantities of the
solutions through the use of the conterterm method inspired by the
AdS/CFT correspondence.

It is worth comparing the solutions obtained here to the
electrically charged rotating dilaton black string solutions
presented in \cite{shey1}. In the present work I have studied the
magnetic spinning dilaton string which produces a longitudinal
magnetic field in AdS spaces which is the correct one generalizing
of the magnetic string solution of Dias and Lemos in dilaton
theory \cite{Lem2}, while in \cite{shey1} I constructed charged
rotating dilaton black string in AdS spaces which is the
generalization of the charged rotating string solutions of
\cite{Lem3} in dilaton gravity. Although solution (\ref{f(r)}) of
the present paper is similar to Eq. (16) of Ref. \cite{shey1}
(except the sign of $c$) and both solutions represent dilaton
string, however, there are some different between the magnetic
string and the electrically charged
dilaton black string solutions. First, the choice of the metric gauge $%
[g_{tt}\varpropto -\rho ^{2}$ and $(g_{\rho \rho })^{-1}\varpropto
g_{\phi \phi }]$  in the magnetic case which is quite different
from the Schwarzschild gauge $[(g_{\rho \rho })^{-1}\varpropto
g_{tt}$ and $g_{\phi \phi }\varpropto \rho ^{2}] $ proposed in
\cite{shey1}. Second, the electrically charged dilaton black
strings have an essential singularity located at $r=0$ and also
have horizons, while the magnetic strings version presented here
have no curvature singularity and no horizon, but have a conic
geometry. Third, when the rotation parameter is nonzero, the
magnetic string has a net electric charge which is proportional to
the rotation parameter, while charged dilaton black string has
always an electric charge regardless of the rotation parameter.

The generalization of the present work to higher dimensions, that
is the magnetic rotating dilaton branes in AdS spaces with
complete set of rotation parameters and arbitrary dilaton coupling
constant is now under investigation and will be addressed
elsewhere.

\acknowledgments{This work has been supported financially by the
Research Institute for Astronomy and Astrophysics of Maragha,
Iran. The author would like to thank the anonymous referee for
his/her useful comments.}


\end{document}